%% file: mqtunneling.tex
\pdfoutput=1
\documentclass[pra,aps,twocolumn,notitlepage,superscriptaddress,10pt]{revtex4-1}
\usepackage{graphicx,graphicx,amsthm,amsmath,amssymb}
\usepackage{wasysym}
\usepackage{color}
\usepackage{bbold}
\usepackage[caption=false]{subfig}
\usepackage{hyperref}

\usepackage{geometry}
\def\com#1{}

\def\ket#1{| #1 \rangle}
\def\bra#1{\langle #1 |}
\def\L{{\rm L}}
\def\R{{\rm R}}

\begin{document}

\title{Computational Role of Multiqubit Tunneling in a Quantum Annealer}

\author{Sergio Boixo\textsuperscript{1}, Vadim N. Smelyanskiy\textsuperscript{1,2}, Alireza Shabani\textsuperscript{1},\\ Sergei V. Isakov\textsuperscript{1}, Mark Dykman\textsuperscript{3}, Vasil S. Denchev\textsuperscript{1},\\ Mohammad Amin\textsuperscript{4}, Anatoly Smirnov\textsuperscript{4}, Masoud Mohseni\textsuperscript{1},\\ Hartmut Neven\textsuperscript{1}\vspace{7mm} \\
   {\normalsize \textsuperscript{1}Google, Venice, CA 90291, USA} \\
{\normalsize \textsuperscript{2}NASA Ames Research Center, Moffett Field, CA 94035, USA}\\
{\normalsize \textsuperscript{3}Department of Physics and Astronomy, Michigan State University,}\\ \normalsize{East Lansing, MI 48824, USA }\\
{\normalsize \textsuperscript{4}D-Wave Systems Inc., Burnaby, BC V5C 6G9, Canada}
}

\begin{abstract}
  \input{abstractsci}
\end{abstract}

\maketitle

\input{main}

\vspace{0.5in} 
\input{acknowledgments}

\bibliography{experimental_tunneling}

\newpage

\end{document}

%% file: abstractsci.tex


  Quantum tunneling, a phenomenon in which a quantum state traverses
  energy barriers above the energy of the state itself, has been
  hypothesized as an advantageous physical resource for
  optimization. Here we show that multiqubit tunneling plays a
  computational role in a currently available, albeit noisy,
  programmable quantum annealer. We develop a non-perturbative theory
  of open quantum dynamics under realistic noise characteristics
  predicting the rate of many-body dissipative quantum tunneling. We
  devise a computational primitive with 16 qubits where quantum
  evolutions enable tunneling to the global minimum while the
  corresponding classical paths are trapped in a false
  minimum. Furthermore, we experimentally demonstrate that quantum
  tunneling can outperform thermal hopping along classical paths for
  problems with up to 200 qubits containing the computational
  primitive. Our results indicate that many-body quantum phenomena
  could be used for finding better solutions to hard optimization
  problems.

%% file: main.tex
Quantum tunneling was discovered in the late 1920s to explain
radioactive decay and field electron emission in vacuum
tubes. Tunneling plays a key role in Josephson junctions, molecular
nanomagnets, and in charge and energy transport in biological and
chemical processes~\cite{mohseni2014quantum}. It is at the core of
many essential technological innovations such as flash memories and
the scanning tunneling microscope. Quantum tunneling has also been
hypothesized as an advantageous mechanism for quantum optimization, in
particular in systems with thin but high energy
barriers~\cite{ray_sherrington-kirkpatrick_1989,Finnila_1994,
  Nishimori_1998,Brooks_99,farhi_quantum_2002,Santoro_2002}.  In
classical cooling optimization algorithms such as simulated annealing
the initial temperature must be high in order to overcome tall energy
barriers. As the algorithm progresses the temperature is gradually
lowered to distinguish between local minima with small energy
differences. This causes the stochastic process to freeze once the
thermal energy is lower than the height of the barriers surrounding
the state.  In contrast, quantum tunneling transitions are still
present even at zero temperature. Therefore, for some energy
landscapes, one might expect that quantum dynamical evolutions can
converge to the global minimum faster than the corresponding classical
cooling process.

Quantum annealing~\cite{Finnila_1994,Nishimori_1998} is a technique
inspired by classical simulated annealing which aims to take advantage
of quantum tunneling. The performance of chips designed to implement
quantum annealing using superconducting electronics has been studied
in a number of recent
works~\cite{mooij1999josephson,harris_experimental_2010b,lanting_cotunneling_2010,johnson2011quantum,boixo2013experimental,dickson2013thermally,mcgeoch2013experimental,
  boixo_evidence_2014,lanting2014entanglement,santra2014max,ronnow_defining_2014,vinci2014hearing,shin2014quantum,vinci2014distinguishing,venturelli2014quantum,albash2014reexamining}. Here
we consider chips manufactured by D-Wave Systems, described in detail
in~\cite{boixo2014computational}.  The qubits are subject to complex
interactions with the environment. We show that even under such
conditions the device performance benefits from multiqubit tunneling.
We consider a computational primitive, the simplest non-convex
optimization problem consisting of just one global minimum and one
false (local) minimum. Quantum evolutions enable tunneling to the
global minimum while the corresponding classical paths are trapped in
the false minimum. A detailed multiqubit master equation accurately
describes the experimental data from the D-Wave Two processor at NASA
Ames. 
We study the temperature dependence of the probability of success for
our computational problem. Consistent with our quantum models, we
experimentally determine that the temperature dependence of the
success probability of the D-Wave chip is opposite to the temperature dependence
predicted by models based on classical paths with thermal hopping.

The goal of quantum annealing is to find low energy states of a ``problem Hamiltonian'' 
\begin{equation}
  H_{\rm P} = - \sum_\mu h_\mu \sigma_\mu^z - \sum_{\mu\nu} J_{\mu\nu} \sigma_\mu^z \sigma_\nu^z\;,
\end{equation}
where the Pauli matrices $\sigma_\mu^z$ correspond to spin variables
with values $\{\pm 1\}$. The local fields $\{h_\mu\}$ and couplings
$\{J_{\mu\nu}\}$ define the problem instance.  Quantum annealing
is charaterized by evolution under the Hamiltonian
\begin{equation}
  H_0(s) = A(s)H_{\rm D} + B(s) H_{\rm P}\;,
\end{equation}
where $H_D=-\sum_\mu\sigma_\mu^x$. The annealing parameter $s$ slowly
increases from 0 to 1 throughout the annealing time $t_{qa}$. Initially
$A(0)\gg B(0)$. With increasing $s$, $A(s)$ monotonically decreases to
0 for $s=1$, whereas $B(s)$ increases.

\begin{figure}[ht]
  \centering
  \includegraphics[width=.7\columnwidth]{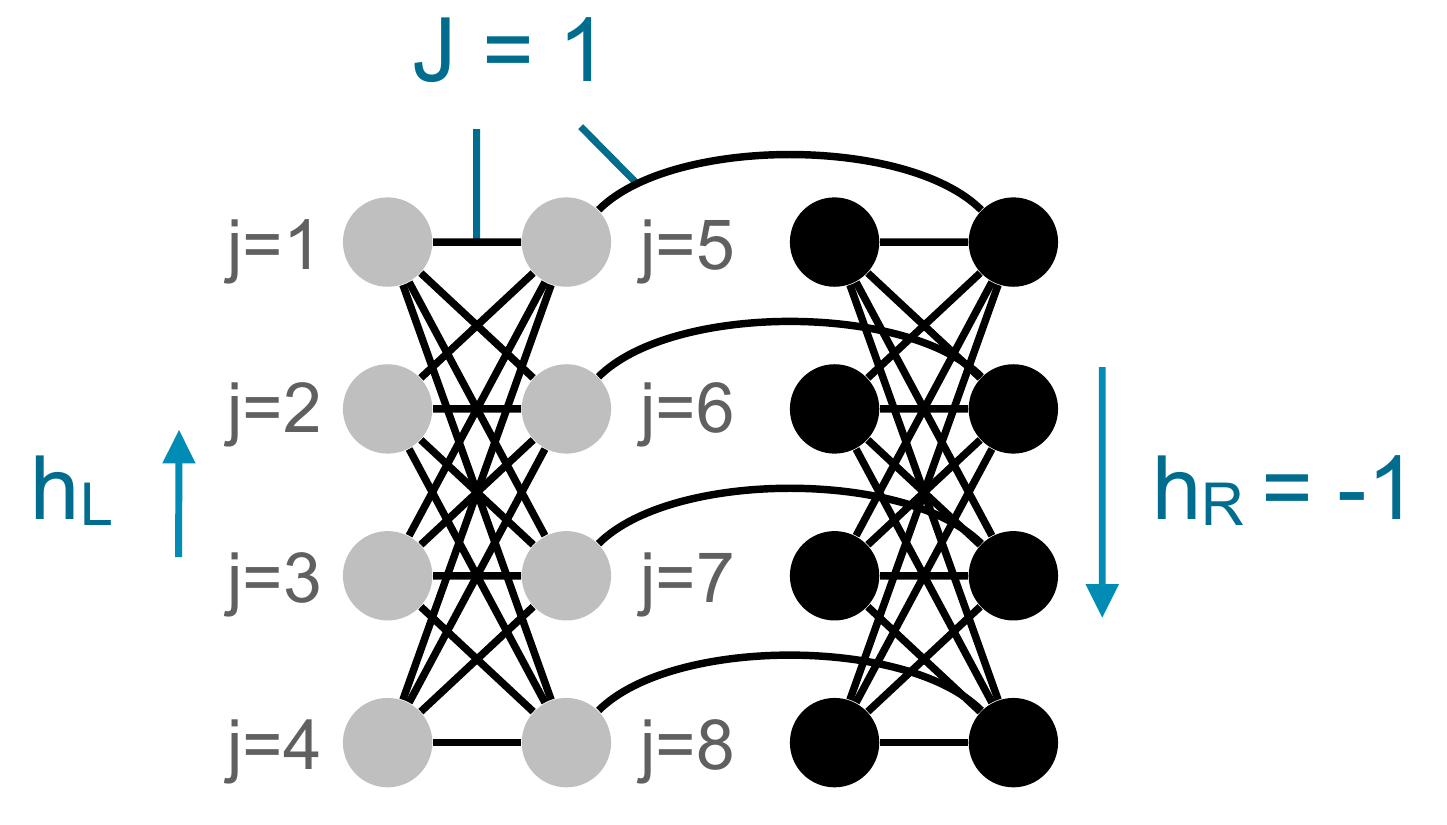}\label{fig:Strong_Weak_Clusters}
  \caption{Graph of the problem Hamiltonian with 16 qubits
    coupled ferromagnetically with J=1 (lines). The applied fields are
    $0 < h_\L < J/2$ ($h_\R=-1$) for the left (right) qubit cell. The
    symmetry and strong intra-cell ferromagnetic coupling makes each
    8-qubit cluster evolve together.}
\end{figure}

The problem Hamiltonian encoding the computational primitive with one
global minimum and one false minimum is depicted in
Fig.~\ref{fig:Strong_Weak_Clusters}.  It consists of two qubit cells,
left and right, each with $n=8$ qubits.  The local fields $0<h_\L<0.5$ and
$h_\R=-1$ are equal for all the spins within each cell, and all the
couplings $J=1$ are ferromagnetic.  The spins within each cell tend to
move together as clusters due to symmetry and the strong
intra-cell ferromagnetic coupling energy.  We choose $|h_\R| >
|h_\L|$, so that in the low energy states of $H_P$ the right cluster
is pointing along its own local field as seen in
Fig.~\ref{fig:Strong_Weak_Clusters}. The difference in energy of the
states with opposite polarization in the left cluster is $n(J-2
h_\L)$. Choosing $h_\L< J/2=0.5$, the global minimum corresponds to both
clusters having the same orientation, while in the false minimum
they have opposite orientations.

Our aim is to distinguish quantum tunneling from thermal activation
along classical paths of product states (which preclude multiqubit
tunneling). We now explain why a classical path continuously connects
the initial global minimum to the final false minimum. At the
beginning of the annealing process $B(s)/A(s) \ll 1 $, and we have
$\langle \sigma^z_\mu\rangle \simeq h_k\, B(s)/A(s)$ (because the
coupling terms are quadratic in the z-polarizations $\langle
\sigma^z_\mu\rangle$). As $h_\L$ and $h_\R$ have opposite signs, so
will the z-projections of spins in the two clusters early in the
evolution.  To escape this path classically all spins in the left
cluster must flip sign, which requires traversing an energy
barrier. The barrier peak corresponds to zero total z-polarization of
the left cluster. Therefore, the barrier grows with the ferromagnetic
energy of the cluster $(n/2)^2 J$. The barrier height is much greater
than the residual energy which grows with $n(J-2 h_\L)$.
 
\begin{figure}[ht]
  \centering
  \includegraphics[width=\columnwidth]{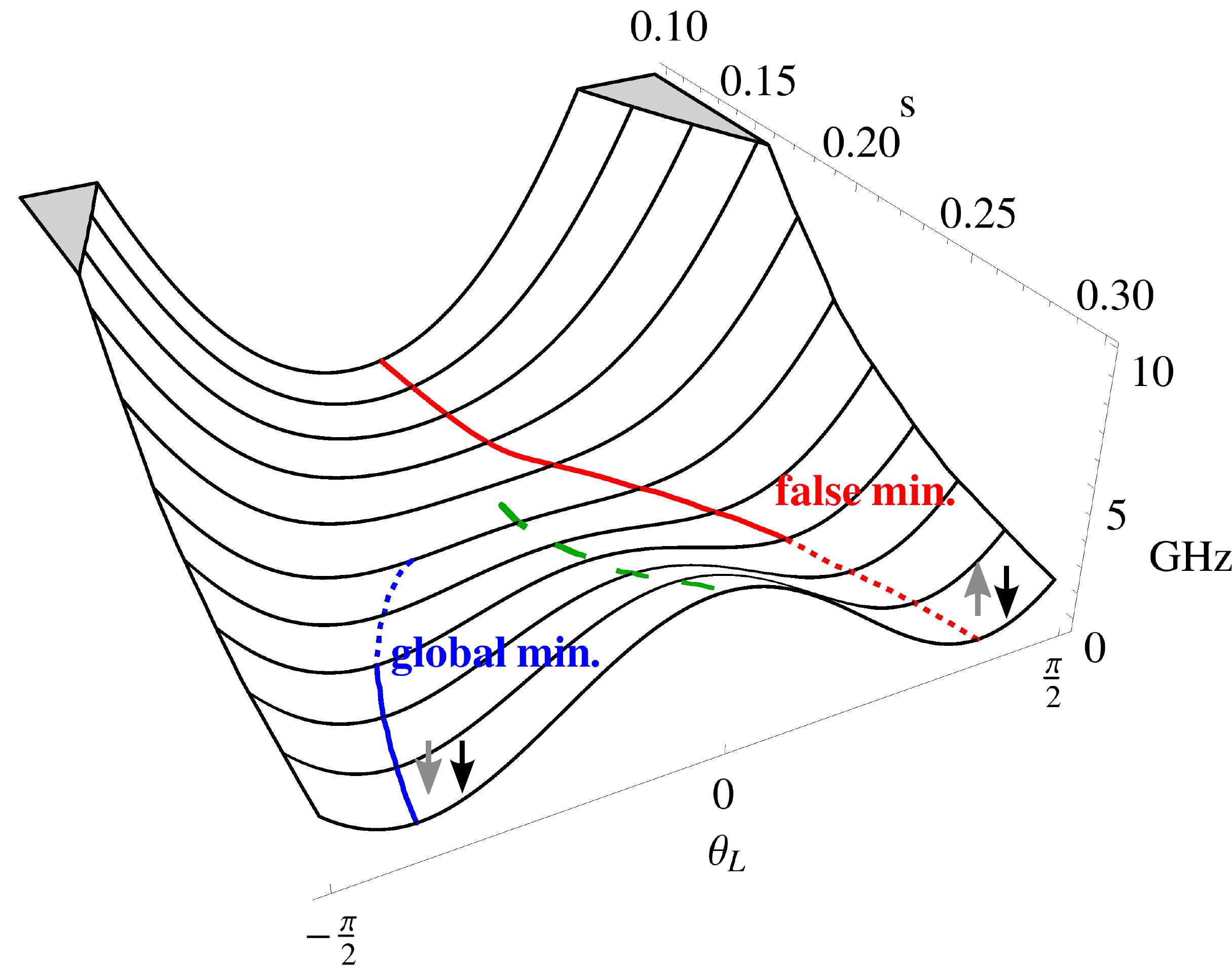}
  \caption{Energy potential using
    $h_\L=0.44$, plotted versus annealing $s$ and tilt angle
    $\theta_\L$ of each spin vector in the left cluster.  The red line
    corresponds to a path that starts in the initial global minimum
    and follows the instantaneous local energy minimum. A second local
    minimum (dashed blue line) forms at the bifurcation point
    $s=0.18$. The global minimum is found in this second path after
    $s=0.24$ (dashed to continuous blue line). To reach this global
    minimum the system state has to traverse the energy barrier
    between them (dashed green line), either by thermal activation or
    by quantum tunneling.}
  \label{fig:contour_3D}
\end{figure}

In order to give a more precise description of the classical paths of
product states, let each qubit be represented by a spin vector in the
$xz$-plane. Denote by $\theta_\mu$ the angle of the spin vector for
qubit $\mu$ with the $x$ quantization axis. We can gain an intuitive
understanding of the effective energy landscape if we assume that all
the qubits in the left (right) cluster have the same angle $\theta_\L$
($\theta_\R$). This assumption is based on symmetry and the strong
intra-cluster ferromagnetic energy. The resulting energy potential can
be derived using more formal methods, like the Villain
representation~\cite{boulatov_quantum_2003}. Figure~\ref{fig:contour_3D}
plots the effective energy potential for the left cluster as a
function of $\theta_\L$ with $h_\L = 0.44$. The classical path (red
line) which follows the local minimum of this effective energy
potential gets trapped in a false minimum and fails to solve the
corresponding optimization problem, as explained in the previous
paragraph.

\begin{figure*}[ht]
  \centering
\subfloat[Probability vs. temperature.]{\includegraphics[width=.4\textwidth]{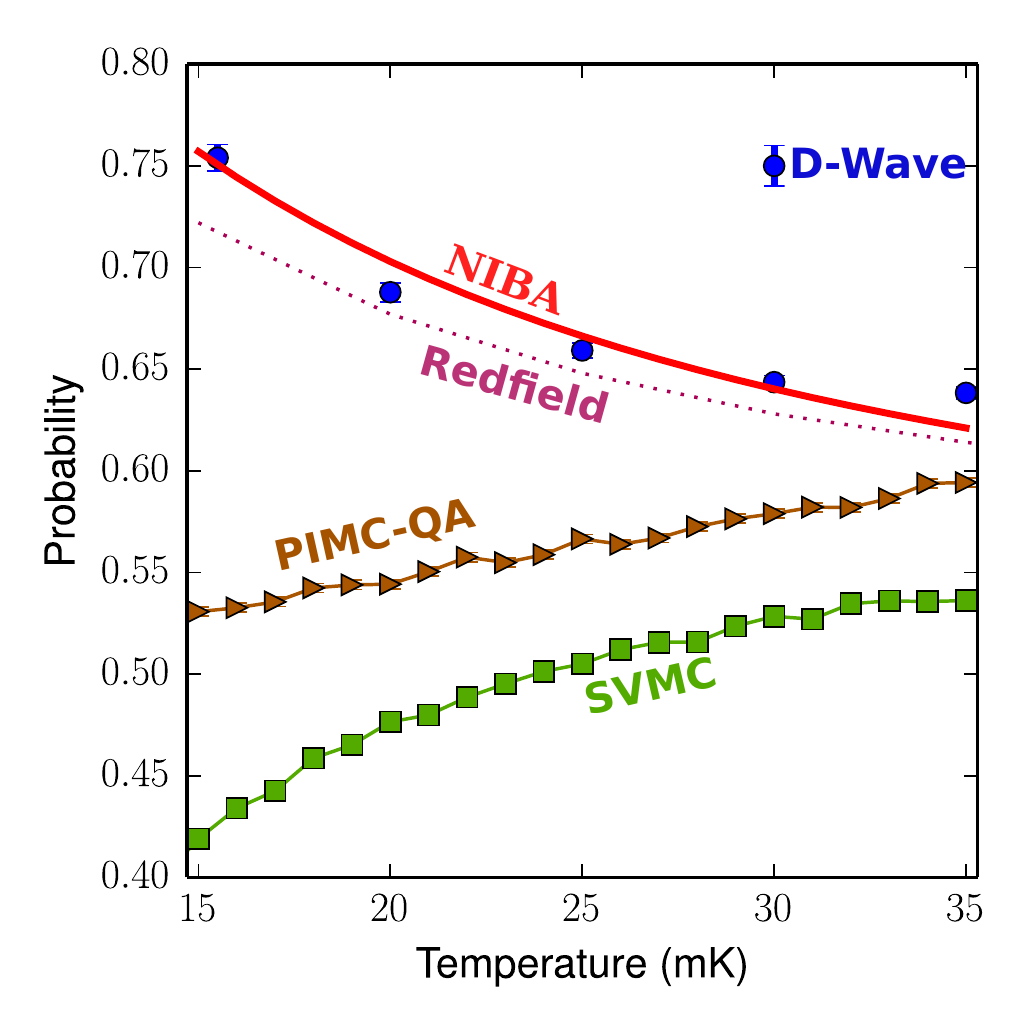}\label{fig:p_vs_t}}
   \subfloat[Probability vs. $h_\L$.]{\includegraphics[width=.4\textwidth]{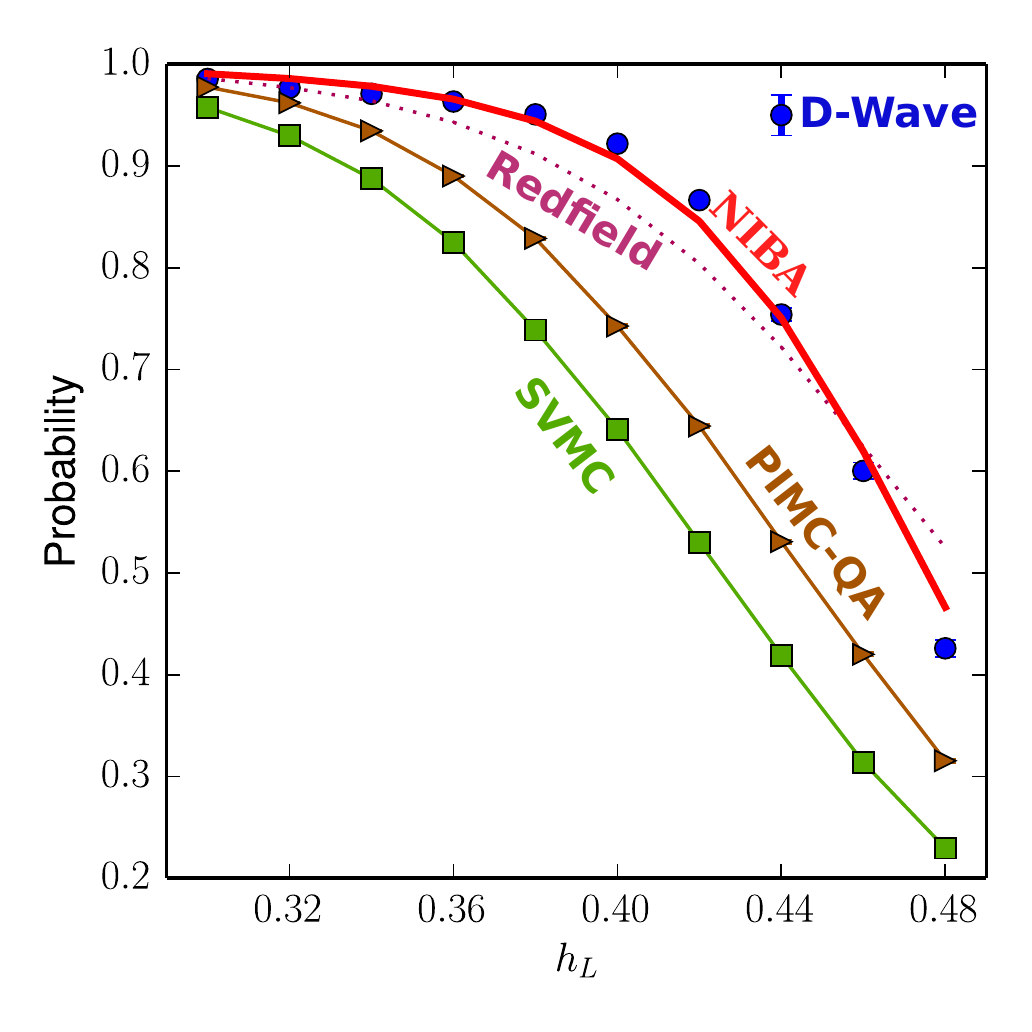}\label{fig:p_vs_h1}}
  \caption{Plots (a)
    and (b) show the probability of success for D-Wave (blue points)
    versus $h_\L$ and temperature. The decreasing probability with
    increasing temperature using $h_L=0.44$ is only matched with
    theories based on quantum tunneling. This is the opposite tendency
    to thermal activation (SVMC). Both Redfield and NIBA use only
    measured parameters (no fitting).  In this temperature range the
    lowest two states (the double well potential) account for all the
    probability (0.9998 for D-Wave, 0.99998 for SVMC).}
\end{figure*}

In the absence of quantum tunneling, the global minimum could be
reached through thermal excitations along classical paths for
over-the-barrier escape from the false minimum. This thermal
activation results in an increasing probability of success with rising
temperature. This intuition is supported by spin vector Monte Carlo
(SVMC), a numerical algorithm consisting in thermal Metropolis updates
of the spin vectors~\cite{shin2014quantum}. Figure~\ref{fig:p_vs_t}
confirms the thermal activation in SVMC. This is opposite to both open
quantum system theory and experiments with the D-Wave chip, which show
a reduction of the probability of success with rising temperature, as
explained later. Furthermore, Figure~\ref{fig:p_vs_h1} shows that the
probability of success for SVMC is lower than the probability of
success for D-Wave and open system quantum models.


\begin{figure}[ht] \centering
  \includegraphics[width=\columnwidth]{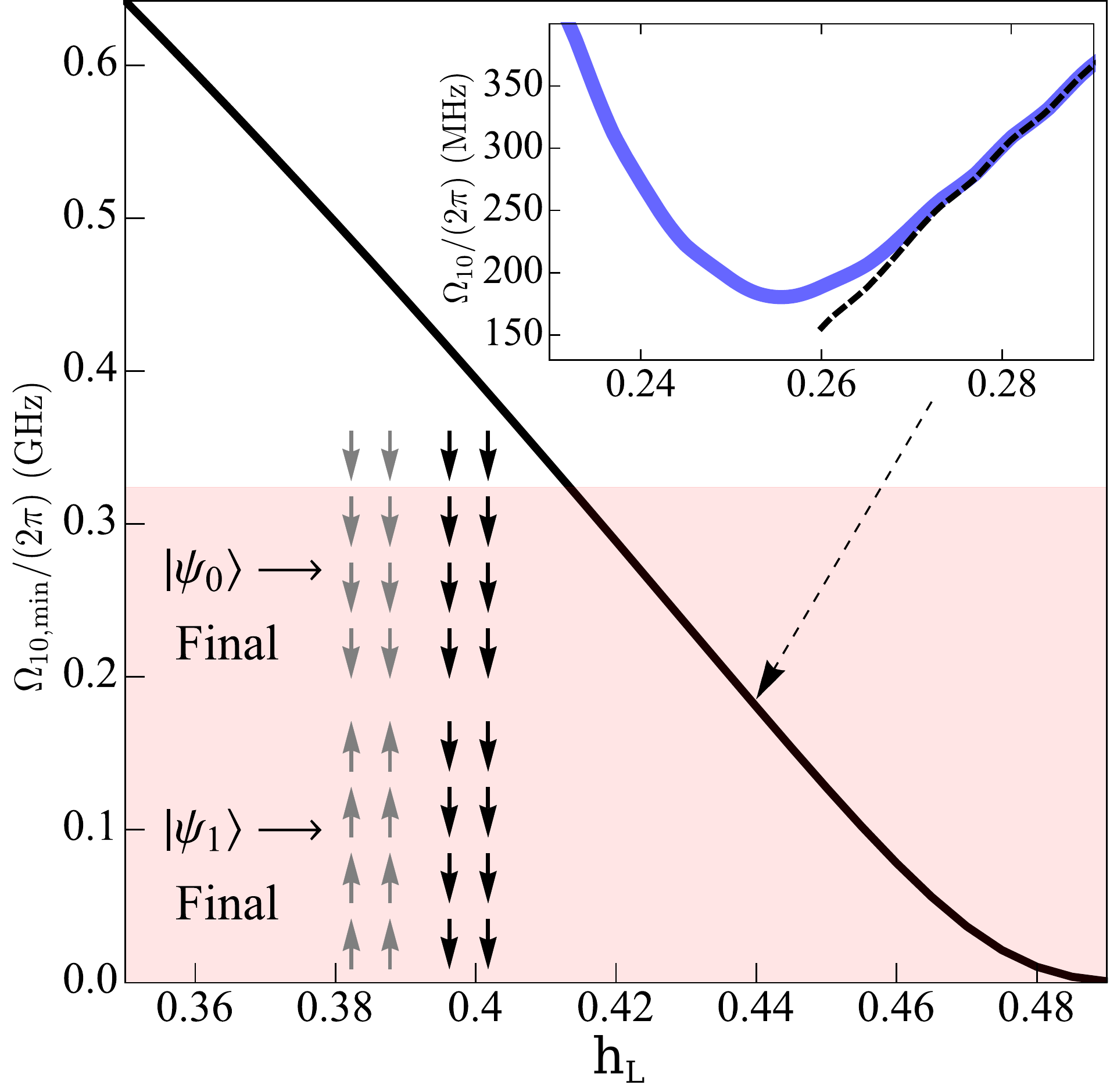}
  \caption{Inset shows the quantum energy gap $\hbar \Omega_{10} =
E_{1}(s)-E_0(s)$ versus $s$, using $h_\L=0.44$. The dashed line is the
gap in the diabatic (pointer) basis. In the main plot, the minimum gap
decreases with $h_\L$. The horizontal boundary of the red-filled area
(324~MHz) corresponds to 15.5~mK, the lowest temperature in our
experiments. The lower inset shows the spin configurations of the two
lowest eigenstates at the end of the annealing.}
  \label{fig:mingaps}
\end{figure}

Quantum mechanically, the system evolution goes through an
``avoided-crossing'' where the two lowest eigenstates $E_1(s)$ and
$E_0(s)$ approach closely to, and then repel from, each other (see
inset in Fig.~\ref{fig:mingaps}). Higher energy states remain well
separated during the evolution. This level repulsion occurs due to the
collective tunneling of qubits in the left cluster between the
opposite $z$-polarizations.  At the point where the gap
$\hbar \Omega_{10}(s)$=$E_1(s)$-$E_0(s)$ reaches its minimum the corresponding adiabatic
eigenstates are formed by the symmetric and anti-symmetric
superpositions of the cluster orientations. The size of the minimum
gap is varies with $h_\L$, as seen in Fig.~\ref{fig:mingaps}.


Under realistic conditions, a quantum annealer can be strongly
influenced by coupling to the environment, for which we introduce a
detailed {\it phenomenological} open quantum system model. We shall
assume that each flux qubit is coupled to its own environment with an
independent noise source; this is consistent with experimental
data~\cite{lanting_cotunneling_2010}.  The coupling of the environment
to each flux qubit is through flux fluctuations, and is proportional to a
$\sigma^z$ qubit operator.  The properties of the noise are determined
by the noise spectral density $S(\omega)$, which is characterized by
single-qubit macroscopic resonant tunneling (MRT) experiments in a
broad range of biases (0.4 MHz $-$ 4 GHz) and temperatures (21~mK $-$
38~mK) for tunneling amplitudes of a single flux qubit below 1 MHz .
The MRT data collected is surprisingly
well-described~\cite{harris_experimental_MRT_2008,PhysRevB.83.180502}
by a phenomenological ``hybrid'' thermal noise model $S(\omega)=S_{\rm
  lf}(\omega)+S_{\rm oh}(\omega)$. Here $S_{\rm oh}(\omega)=\hbar^2 \eta
\omega e^{-\omega\tau_c}/(1-e^{-\hbar\omega/k_B T})$ denotes the
high-frequency part, and has Ohmic form with dimensionless coupling
$\eta$ and cutoff frequency $1/\tau_c$ (assumed to be very large).
The low-frequency part $S_{\rm lf}$ is of the $1/f$
type~\cite{Martinis:2003} and in current D-Wave chips
this noise is coupled to the flux qubit relatively strongly. Its
effect can be described with only two parameters: the width $W$ and the
Stokes shift $\epsilon_p$ of the MRT
line~\cite{amin_macroscopic_2008}. The experimental shift value is
related to the width by the fluctuation-dissipation theorem
($\epsilon_p = \hbar W^2/ 2 k_B T$) and represents the reorganization
energy of the environment.  The values of the noise parameters
measured at the end of the annealing ($s=1$) for the D-Wave Two chip
are $W/(2\pi)=0.40(1)\,{\rm GHz}$ and $\eta=0.24(3)$.

In the analysis of the transitions between the states we start from
the initial (gapped) stage when the  instantaneous  energy gap $\hbar \Omega_{10}(s)$ between the two lowest eigenstates
$|\psi_0(s)\rangle$, $|\psi_1(s)\rangle$    is sufficiently
large compared to the linewidth $\hbar W$. Then the coupling to the environment can be treated as a
perturbation and the transition rate between these states  is then given by
Fermi's golden rule 
  $\Gamma_{1\rightarrow 0}(s) \approx  \mathbb a(s) S(\Omega_{10}(s))/\hbar^2$. 
Here
\begin{equation}
\mathbb a(s) = \sum_{\mu=1}^{2n}|\bra{\psi_0(s)} \sigma_\mu^z \ket{\psi_1(s)}|^2
\end{equation}
is a  sum of (squared)  transition matrix elements between  the two eigenstates.   

\begin{figure}[ht]
  \centering
  \includegraphics[width=\columnwidth]{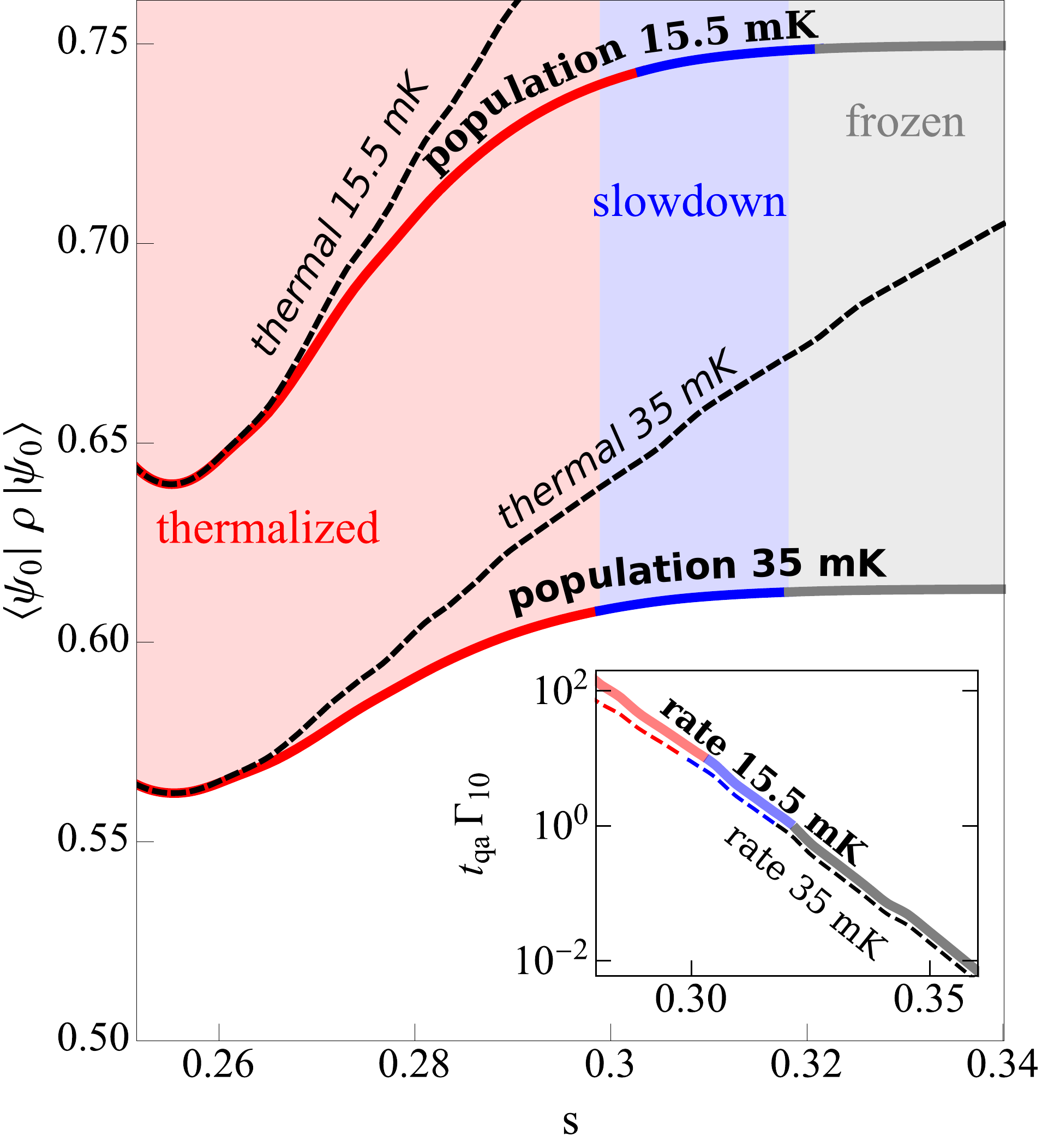}
  \caption{Solid lines correspond to the modeled population of
    the lowest energy eigenstate along the quantum annealing process
    using $h_L=0.44$ at 15.5~mK (top line) and 35~mK (bottom
    line). Dashed lines correspond to the thermal equilibrium
    population.  In the thermalization phase (red) the transition rate
    is fast and the population remains close to thermal
    equilibrium. As the multiqubit energy barrier increases, the
    transition rates are exponentially reduced with $s$, as shown in
    the inset. We define the slowdown regime (blue) as $t_{qa}
    \Gamma_{1\rightarrow 0} < 10$, and the frozen regime (gray) as $t_{qa}
    \Gamma_{1\rightarrow 0} < 0.1$. Comparing the data at 15.5 and 35~mK, we see a
    small change in the transition rate relative to the larger change
    in the thermal equilibrium ground state population. Therefore, the
    probability of success is lower at higher temperature.}
  \label{fig:freez}
\end{figure}

In the minimum gap region the (squared) matrix element $\mathbb a(s)$
for the transition rate is large, and the system is thermalized (see
Fig.~\ref{fig:freez}). More precisely, we have $\Gamma_{1\rightarrow
  0} \gg 1/t_{qa}$, where the inverse of the annealing time $1/t_{qa}$
is an approximation for the annealing rate. The ground state
population is given by the Boltzmann distribution at the experimental
temperature.

After the avoided-crossing region we observe a steep exponential
fall-off of the matrix element $\mathbb a(s)$ with $s$, eventually
causing \emph{multiqubit freezing} (see Fig.~\ref{fig:freez}).
Multiqubit freezing is quite distinct from single qubit
freezing. Single qubit tunneling~\cite{johnson2011quantum} decays
slowly as the magnitude of the transverse field $A(s)$ decreases. The
multiqubit transition rate, however, decays exponentially fast (see
inset of Fig.~\ref{fig:freez}). This is due to the increasing
effective barrier width (see Fig.~\ref{fig:contour_3D}), which results
in an exponential decrease of quantum tunneling and in a {\it
  slowdown} of the transition rate $\Gamma_{1\rightarrow
  0}$. Formally, the barrier width corresponds to the Hamming distance
\begin{equation}
\mathbb h(s)=\sum_{\mu=1}^{2n}|\bra{\psi_0} \sigma_\mu^z\ket{\psi_0}-\bra{\psi_1} \sigma_\mu^z \ket{\psi_1}|^2/4
\end{equation}
between the
opposite $z$-orientations of the left cluster in the two lowest energy
eigenstates. The exponential sensitivity of multiqubit tunneling to
the width or Hamming distance $\mathbb h(s)$ is the cause of the
exponential decay of the matrix element $\mathbb a(s)$, and of the
multiqubit freezing.

We distinguish a slowdown phase (roughly $0.1 < t_{qa} \Gamma_{1\rightarrow 0} <
10$) and a {\it frozen} phase ($t_{qa} \Gamma_{1\rightarrow 0} < 0.1$). In the
frozen phase, there are no dynamics. Part of the system population
remains trapped in the excited state $\ket{\psi_{1}(s)}$ corresponding
to the false minimum of the effective potential until the end of the
quantum annealing process (see Fig.~\ref{fig:freez}).

The success probability of quantum annealing is (roughly) determined
by the thermal equilibrium ground state population during the slowdown
phase. When the temperature grows, the ground state population
decreases appreciably, while the transition rate changes little (see
Fig.~\ref{fig:freez}). This results in the observed \emph{thermal
  reduction} (see Fig.~\ref{fig:p_vs_t}). 

When the energy gap is similar to (or smaller than) the noise
linewidth $W$ the environment cannot be treated as a perturbation. We
develop a multiqubit non-perturbative analysis in the spirit
of 
the Non-interacting Blip Approximation (NIBA)~\cite{Leggett:1987} that
covers all QA stages. In the slowdown phase, when the Hamming distance
approaches its maximum value $\mathbb h \sim n$, the instantaneous
decay rate of the first excited state takes the form
\begin{widetext}
\begin{align}
  \Gamma_{1\rightarrow 0} =\int_{-\infty}^{\infty}d\tau\,e^{i \Omega_{10}\tau - \mathbb h (i \epsilon_p \tau + {(W
      \tau)^2}/{2})} \left [\frac{\pi \tau_c}{i\beta} {\rm csch} 
    \frac{(\tau-i \tau_c)}{\beta/\pi}\right]^{\frac{ \mathbb h \eta}{2\pi}} D(\tau)\;,
\end{align}
\end{widetext}
where $1/\tau_c$ is the Ohmic noise cutoff
frequency, $\beta$ is $\hbar/k_B T$, and the factor $D(\tau)$ is
provided in~\cite{boixo2014computational}. The dependence on the
annealing parameter $s$ is implicit.  The factor $D(\tau)$ is related
to the tunneling permeability of the potential barrier in
Fig.~\ref{fig:contour_3D} (similar to the coefficient $\mathbb a$).
The above expression describes collective tunneling of the left qubit
cluster assisted by the environment. The crucial difference from the
single qubit MRT
theory~\cite{amin_macroscopic_2008,PhysRevB.83.180502} is that the
parameters of the environment in the transition rate are rescaled by
the barrier width or Hamming distance $\mathbb h(s)$.  The effective
low-frequency noise linewidth is $\mathbb h^{1/2}(s) W(s)$, the
reconfiguration energy is $\mathbb h (s) \epsilon_p(s)$ and the Ohmic
coefficient is $\mathbb h(s) \eta(s)$. This is important at the late
stages of quantum annealing when $\mathbb h \sim n \gg 1$.

We observe a very close correspondence between the results of the
analysis with the NIBA Quantum Master Equation for the dressed cluster
states and the D-Wave Two data displayed in Fig.~\ref{fig:p_vs_h1}.
We emphasize that for NIBA (and the standard Redfield equation with $S_{\rm oh}(\omega)$) we do not have
any parameter fitting: the parameters are obtained from MRT
experiments, as explained above.


Figure~\ref{fig:p_vs_t} shows the success probability, as a function of
temperature, for the D-Wave Two chip, open system quantum simulation and the
classical-path model (SVMC) for $h_\L = 0.44$. The D-Wave Two experimental data
clearly shows thermal reduction: decreasing probability of success
(ground state population) with temperature. This is a consequence of
quantum tunneling, as seen in open quantum system theory. Contrary to
the experimental data, SVMC shows thermal activation, with increasing
success probability for increasing temperature.  For a wide range of
plausible parameters, only the quantum models show thermal reduction
in these instances. The probability of success of SVMC is also lower
than D-Wave Two data at the same temperature.

For $h_\L$ close to the degeneracy value $h_\L=J/2$ the minimum gap
$\Omega_{10}^{\min}$ becomes small, as seen in
Fig.~\ref{fig:mingaps}. Where $\Omega_{10} \ll W$, the adiabatic basis
of the instantaneous mutiqubit states $\{\ket{\psi_0(s)}$,
$\ket{\psi_{1}(s)}\}$ loses its physical significance.  Because the
coupling to the bath is relatively strong here, the system quickly
approaches the states corresponding to predominantly opposite cluster
orientations, similar to diabatic states (see inset of
Fig.~\ref{fig:mingaps}).  Transitions between these states, also
called pointer states~\cite{Zurek81}, occur at a much slower rate as a
consequence of the polaronic effect. As a result, for sufficiently
small mininum gaps the multiqubit freezing starts before the avoided
crossing and the success probability increases with
temperature~\cite{dickson2013thermally}.

\begin{figure}[ht]
  \centering
\includegraphics[width=\columnwidth]{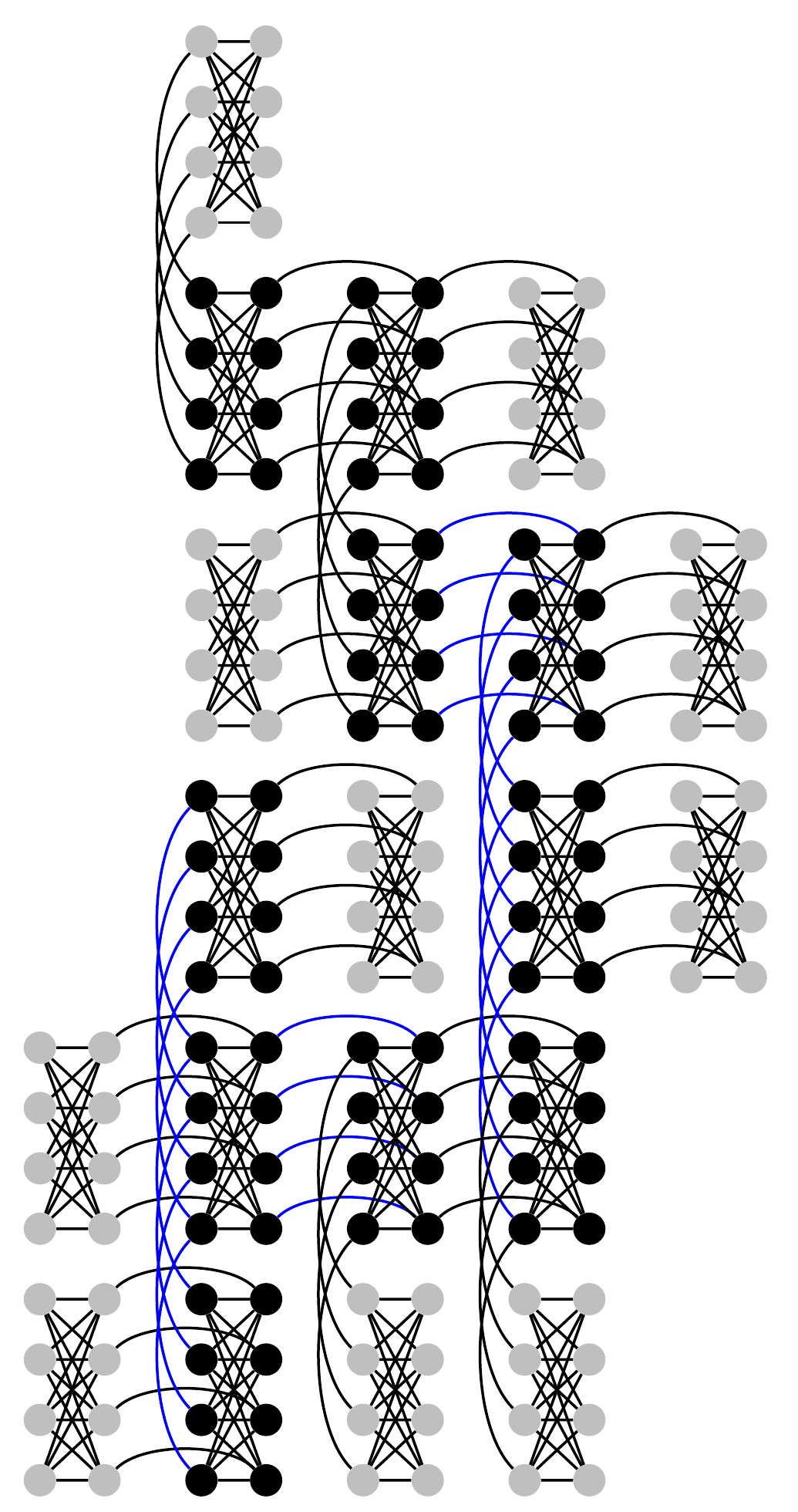}
  \caption{Larger problems that contain the tunneling probe
    ``motif'' as subproblems. As in
    Fig.~\ref{fig:Strong_Weak_Clusters}, the black (grey) cluster has
    a strong $h_R=-1$ (weak $h_L=0.44$) local field. The black
    clusters are connected in a glassy fashion to make the problem
    less regular: all connections between any two neighboring black
    clusters are set randomly to either $-1$ or $+1$. The $-1$
    connections are depicted in blue. }
  \label{fig:size5_problem}
\end{figure}

\begin{figure}[ht]
  \centering
\includegraphics[width=\columnwidth]{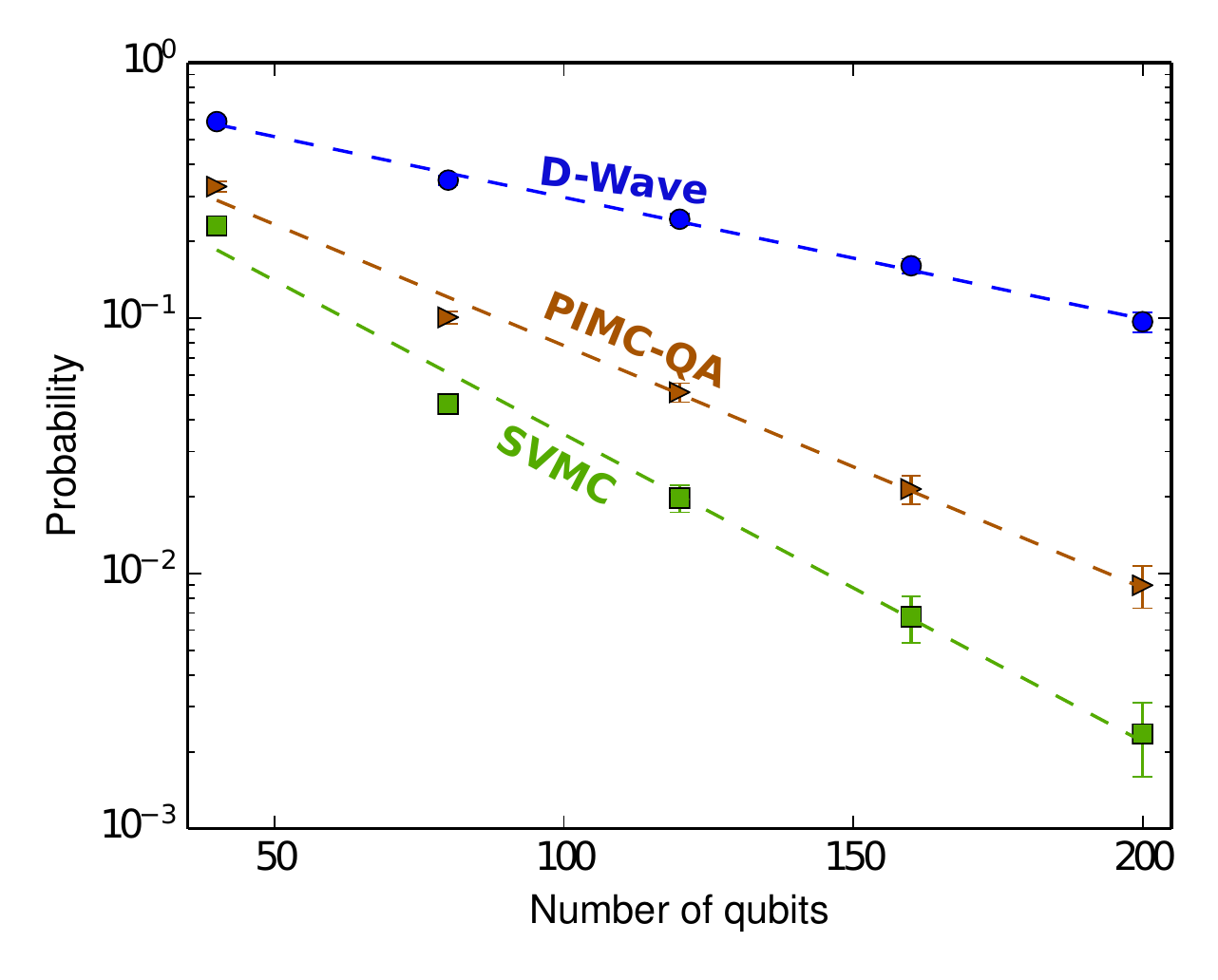}
  \caption{Success probability for a
    glass of clusters as a function of the number of qubits
    involved. We fit the mean probability of success $p(n_q) \propto
    \exp(- \alpha n_q)$ as a function of the number of qubits $n_q$ (dashed lines).
    The fitting exponent $\alpha$ for the D-Wave Two data is $(1.1 \pm
    0.05)\cdot 10^{-2}$, while the fitting exponent for PIMC-QA is
    $(2.2 \pm 0.1) \cdot 10^{-2}$ and for the SVMC numerics is $(2.8
    \pm 0.17)\cdot 10^{-2}$. The error estimates for the exponents are
    obtained by bootstrapping. }
  \label{fig:clusters_glass}
\end{figure}

A generalization of the 16 qubit problem to a larger number of qubits
is achieved by studying problems that contain the same ``motif''
multiple times within the connectivity graph, see
Fig.~\ref{fig:size5_problem}.  The success probabilities for up to 200
qubits are shown in Fig.~\ref{fig:clusters_glass}. We fit the average
success probability as $p(n_q) \propto \exp(- \alpha n_q)$, where
$n_q$ is the number of qubits. The fitting exponent $\alpha$ for the
D-Wave Two data is $(1.1 \pm 0.05)\cdot 10^{-2}$, while the fitting
exponent for the SVMC numerics is $(2.8 \pm 0.17)\cdot 10^{-2}$. We
conclude that, for instances with multiqubit quantum tunneling, the
D-Wave Two processor returns the solution that minimizes the energy
with consistently higher probability than physically plausible models
of the hardware that only employ product states and do not allow for
multiqubit tunneling transitions.


The correlation between D-Wave's experimental data and Path Integral
Monte Carlo along the Quantum Annealing schedule (PIMC-QA) has been
studied in recent
works~\cite{boixo_evidence_2014,ronnow_defining_2014,albash2014reexamining}.
For completeness, we study PIMC-QA using similar parameters as
in~\cite{boixo_evidence_2014}. PIMC-QA gives a probability of success
between SVMC and the quantum models (Figs.~\ref{fig:p_vs_h1} and
\ref{fig:clusters_glass}), and does not show thermal reduction for
$h_\L = 0.44$ (Fig.~\ref{fig:p_vs_t}).

A way to think of multiqubit tunneling as a computational resource is
to regard it as a form of large neighborhood search. Collective
tunneling transitions involving $K$ qubits explore a $K$ variable
neighborhood, and there is a combinatorial number of such
neighborhoods. We find that the current generation D-Wave Two annealer
enables tunneling transitions involving at least 8 qubits. It will be
an important future task to determine the maximal $K$ attainable by
current technology and how large it can be made in next generations.
The larger $K$, the easier it should be to translate the quantum
resource ``$K$-qubit tunneling'' into a possible computational
speedup. We want to emphasize that this paper does not claim to have
established a quantum speedup. To this end one would have to
demonstrate that no known classical algorithm finds the optimal
solution as fast as the quantum process.  To establish such an
advantage it will be important to study to what degree collective
tunneling can be emulated in classical algorithms such as Quantum
Monte Carlo or by employing cluster update methods. However, the
collective tunneling phenomena demonstrated here present an important
step towards what we would like to call a {\it physical speedup}: a
speedup relative to a hypothetical version of the hardware operated
under the laws of classical physics.

%% file: acknowledgments.tex
{\small \noindent \textbf{Acknowledgments}\\ 
\noindent 
We would like to thank
John Martinis, Edward Farhi and Anthonty Leggett for useful discussions  and
reviewing the manuscript. We also thank Ryan Babbush and Bryan O'Gorman for reviewing the manuscript, and Damian Steiger, Daniel Lidar and Tameem Albash for comments about the temperature experiment. The work of V.N.S. was
supported in part by the Office of the Director of National
Intelligence (ODNI), Intelligence Advanced Research Projects Activity
(IARPA), via IAA 145483 and by the  AFRL Information Directorate under grant F4HBKC4162G001.}
\\